\documentclass[12pt]{article}

\usepackage[pdftex]{graphicx} \DeclareGraphicsExtensions{.pdf,.png,.jpg,.jpeg,.mps}
\usepackage{subfigure}
\DeclareGraphicsExtensions{.pdf}
\begin{document}

%\bibliographystyle{naturemag}
%\bibliographystyle{natbib}

% Set page numbering to Arabic numerals
\pagenumbering{arabic}
\newpage

\title{\LARGE
Nonparametric survival analysis and vaccine efficacy using Dempster-Shafer analysis\\
\bigskip}
\author{}
\date{}
\maketitle

\thispagestyle{empty}

\begin{center}
\vspace{1in}
\thispagestyle{empty}

Paul T. Edlefsen$^1$ and Arthur P. Dempster$^2$\\

\vspace{.2in}

\today
\vspace{.6in}

$^1$Statistical Center for HIV/AIDS Research and Prevention\\
Vaccine and Infectious Disease Division\\
Fred Hutchinson Cancer Research Center\\

$^2$Harvard University Statistics Department \\

\end{center}
\clearpage

%%%%%%%%%%%%%%%%%%%%%%%%%%%%%%%%%%%%%%%%%%

%% \clearpage
%% \tableofcontents
%% \clearpage
%% \listoftables
%% \clearpage
%% \listoffigures
%% \clearpage

\section{Abstract}
We introduce an extension of nonparametric DS inference for arbitrary univariate CDFs to the case in which some failure times are (right)-censored, and then apply this to the problem of assessing evidence regarding assertions about relative risks across two populations.  The approach enables exploration of the sensitivity of survival analyses to assumed independence of the missing data process and the failure proces.  We present an application to the partially efficacious RV144 (HIV-1) vaccine trial, and show that the strength of conclusions of vaccine efficacy depend on assumptions about the maximum failure rates of the subjects lost-to-followup.

\section{Introduction}
Given data from an HIV vaccine trial, we are interested in determining whether the rate of HIV acquisition is lower among vaccine recipients ($r_v$) than among placebo recipients ($r_p$) (over a given period of time-since-vaccination, eg within the first 18 months).  We are furthermore interested in estimating the ratio of these rates (with a ratio of $r_v/r_p < 1$ indicating a lower rate among vaccine recipients).  Finally, we define vaccine efficacy (VE) as $1 - r_v/r_p$, and wish to test assertions about this quantity, such as ``VE $>$ 50\%''.

The data is of the form of right-censored survival data.  That is, we have a total number of subjects $m_p$ assigned to the placebo group and a total number $m_v$ assigned to the vaccine group.  For each subject $i$ we have an estimated infection time $t_i$ if subject $i$ became infected during the course of the trial, or a loss-to-followup time $l_i$ if he did not become infected (this might be the total trial duration for most participants, but could also be some earlier time if the subject dropped out of the trial).  We also know the vaccine treatment assignment indicator $y_i$ for each subject, such that $\sum_i{ y_i } = m_v$ and $\sum_i{ ( 1 - y_i ) } = m_p$.

Wishing to avoid parametric assumptions, we temporarily eschew the usual Cox regression approach, and introduce instead a nonparametric method employing Dempster-Shafer (DS) analysis.  Unlike the alternative nonparametric Kaplan-Meier method, in which right-censored data are accomodated by conditioning on non-censoredness (or ``at risk''-ness), this approach considers the whole dataset for its inferences.  As such its conclusions are not subject to the post-randomization selection bias concerns which limit causal interpretability of estimates based on Kaplan-Meier curves.

Like the Kaplan-Meier approach, this DS method for estimating hazard ratios is based on estimates of the cumulative distribution functions (CDFs) of the ``failure times'' (here infection times) within each treatment group.  If we knew the (true, population) CDFs $F_v$ and $F_p$ for these groups, we could determine the failure rates $r_v$ and $r_p$ for any arbitrary time interval $(t_l, t_u)$ by eg $r_v = F_v( t_u ) - F_v( t_l )$.  This is just the population fraction that fail in that time interval.

To describe the method we therefore begin with the estimation of a single CDF.  Thereafter we will consider how two such estimates may be used in concert to make inferences about a ratio of two rates such as $r_v/r_p$.  We note throughout that estimates of CDFs have many potential uses, of which the present discussion will focus on just this one (which has relevance to the analysis of survival data such as the vaccine trial data which we use as a concrete illustration).

\section{Dempster-Shafer Analysis}
Dempster-Shafer analysis (DSA) is an approach to statistical reasoning which tolerates evidence that is ambiguous with respect to an assertion of interest.  For example, using DSA to address the hypothesis that a possibly-unfair coin has probability $p=.5$ of heads after observing 3 heads in 10 tosses yields evidence of the form $(P,Q,R)$, where $P$ is evidence ``for'' the hypothesis that $p=.5$, $Q$ is evidence ``against'' it, and the remainder, $R = 1 - (P+R)$, is evidence that is ambiguous with respect to the hypothsis (we say it is the probability of ``don't know'').

[etc .. copied and perhaps altered from the DSBanff paper or elsewhere]

\section{Nonparametric DS Estimation of a CDF}
We now turn to the nonparametric DS estimation of a univariate, invertable CDF $F$ from data assumed to have been randomly sampled from it.  Suppose we observe $m$ data points and order them such that $x_1 \leq x_2 \leq \dots \leq x_m$.  We make inferences about $F$ via the auxiliary random variables $Y_1, .., Y_m$, which are independent and Uniformly distributed.  The intuition is as follows: if we knew the inverse function $F^{-1}$, we could have generated the data by first drawing $m$ uniforms and then applying to each the inverse CDF, such that for any $y$ we define a corresponding $x = F^{-1}( y )$.  On an x-y plot of the CDF $F$, we could draw a horizontal line from each $y_i$ to the $F$ curve, and from there drop vertically to the x axis to determine the value of $x_i$.

We needn't assume that the data were in fact generated in this way, but this perspective allows us to make inferences about $F$.  We define random variables $Y_i$ for the height of the CDF at each observed value $x_i$.  Since the $x_i$ are ordered, the correspondingly ordered $Y_i$ are jointly distributed as the order statistics of $m$ uniforms.

Evidence for and against assertions about the function $F$ may be assessed via the random variables $Y$.  For instance if we observe 10 ordered failure times $x_1, x_2, \dots, x_{10}$, and we wish to infer the population fraction whose failure times fall between the fourth and sixth of these (that is, we wish to infer $F^{-1}(x_6) - F^{-1}(x_4)$, we may do so via the joint distribution of the corresonding random variables $Y_6$ and $Y_4$ (as $Y_6 - Y_4$).  That difference is distributed as $(Y_6-Y_4) \sim \hbox{Beta}( 2, 9 )$.

DSA allows us to make inferences about the CDF at points other than the observed values, since we are able to represent our uncertainty using a mass function over sets (or intervals) of values.  For instance, if the values of $x_1, x_2, x_3, \hbox{and} x_4$ (of 10 observed failure times) were $10, 30, 55, \hbox{and} 100$, respectively, and we were interested in testing the hypothesis that at least 15\% of the population fails between times 25 and 75, we could accumulate evidence for this assertion as the probability that $( Y_3-Y_2 ) \geq 15\%$ , and against it by the probability that $(Y_4 - Y_1) < 15\%$.  Such evidence will not generally sum to 1; the remaining evidence is ambiguous and contributes to the residual ``don't know''.  In this instance, we get a $(P,Q,R)$ of $(P=(1 - pbeta( .15, 1, 10 ), Q=pbeta( .15, 3, 8 ), 1-(P+Q)) = (0.20, .18, .62)$.  Note that these events are mutually exclusive and thus may be computed separately.

%% I got that by using:
%% computeEvidenceForIntervalHavingQuantileWidth( dataset.times = c( 10, 30, 55, 100, 105, 106, 107, 108, 109, 110 ), dataset.itt.n = 10, dataset.failures.by.time = cumsum( rep( 1, 10 ) ), dataset.ltfs.by.time = rep( 0, 10 ), assertion.min.quantile.width = .15, assertion.max.quantile.width = 1, assertion.time.low = 25, assertion.time.high = 75, num.samples = 100000 ), and verified it's the same as P = 1 - pbeta( .15, 1, 10 ), Q=pbeta( .15, 3, 8 ).

If we introduce right-censored data of the ``loss to followup'' (LTF) type, we simply treat the number of failures by a particular time (and hence the order-statistic-of-interest) as somewhere between the observed cumulative number of failures and that number plus the cumulative number of LTFs by that time.  For instance if we modify the above example by converting one of the previously-observed-to-be-larger values (eg $x_{10}$) by a subject that is lost to followup at time 50, then we aren't sure whether the order-indices of ``$x_3=55$'' and ``$x_4=100$'' are $(3,4)$, $(3,5)$, or $(4,5)$.  This increases $R$, because now we calculate the evidence $Q$ against the hypothesis as the probability that $Y_5-Y_1 <15\%$, because we don't know whether $x_4$ is above or below the failure time of the lost-to-followup subject.  Thus we get a $(P,Q,R)$ of $(P=(1 - \hbox{pbeta}( .15, 1, 10 ), Q=\hbox{pbeta}( .15, 4, 7 ), 1-(P+Q)) = (.2, .05, .75)$.

%% I got that by using:
%% computeEvidenceForIntervalHavingQuantileWidth( dataset.times = c( 1, 30, 50, 55, 100, 105, 106, 107, 108, 109 ), dataset.itt.n = 10, dataset.failures.by.time = cumsum( c( 1, 1, 0, 1, 1, 1, 1, 1, 1, 1 ) ), dataset.ltfs.by.time = cumsum( c( 0, 0, 1, 0, 0, 0, 0, 0, 0, 0 ) ), assertion.min.quantile.width = .15, assertion.max.quantile.width = 1, assertion.time.low = 25, assertion.time.high = 75, num.samples = 100000 )
%% and verified the new Q as equal to pbeta( .15, 4, 7 ).

Note that we know that the number of new failures between any two time points is bounded below by the observed number of new failures occuring after the lower time point and up to-and-including the upper time point.  It is bounded above by that number plus the total number of LTFs that have accumulated up to the upper time point.  If we represent the data as a two-row matrix $C$ with cumulative sums of failures $C_{1,k}$ and cumulative sums of LTFs $C_{2,k}$ at each discrete time point $t_k$ for which we have any observations, then the number of new failures between any two of these time points $t_j < t_k$ is bounded below by $d_{j,k} := C_{1,k} - C_{1,j}$ and above by $e_{j,k} := d_{j,k} + C_{2,k}$.

Returning to our initial example (with no LTFs), suppose we wish to test the hypothesis that between 10\% and 20\% of the population fails between times 25 and 75. The evidence for this assertion is the probability of the event that both $( Y_3-Y_2 ) \geq 10\%$ AND $(Y_4 - Y_1) \leq 20\%$.  The evidence against the assertion is given by the probability that either $(Y_4 - Y_1) < 10\%$ OR $(Y_3-Y_2) > 20\%$.  In general, to test a hypothesis that between (quantiles) $q_l$ and $q_u$ of the population falls between (times) $t_l$ and $t_u$, we first need to identify the indices into the columns of matrix $C$ of the nearest observed time points to $t_l$ and $t_u$, both above ($k_{t_l}^a$ and $k_{t_u}^a$) and below ($k_{t_l}^b$ and $k_{t_u}^b$) each.  In our example, $t_l = 25$, $t_l = 75$, $k_{t_l}^b = 1$, $k_{t_l}^a = 2$, $k_{t_u}^b = 3$, $k_{t_u}^a = 4$.

With these in hand, the general formula, in the absence of LTFs, is that the evidence for the hypothesis (that between $q_l$ and $q_u$ of the population fails between times $t_l$ and $t_u$) is given by the probability of the event that both $Y_{C_{1,k_{t_u}^b}}  - Y_{C_{1,k_{t_l}^a}} \geq q_l$ AND $Y_{C_{1,k_{t_u}^a}} - Y_{C_{1,k_{t_l}^b}} \leq q_u$.  The evidence against it is given by the sum of the probabilities that $Y_{C_{1,k_{t_u}^a}} - Y_{C_{1,k_{t_l}^b}} < q_l$ and that $k_{t_u}^b  - Y_{C_{1,k_{t_l}^a}} > q_u$.  Due to the symmetry of the intervals separating the uniform order statistics $Y$, the actual orders are irrelevant to these probability calculations: the differences are sufficient.  We will refer to the difference $C_{1,k_{t_u}^b}  - C_{1,k_{t_l}^a}$ as the ``internal interval count'' $v_n$ and to the difference $C_{1,k_{t_u}^a} - C_{1,k_{t_l}^b}$ as the ``external interval count'' $v_x$.  What matters is the probability distributions of intervals $W_n$ and $W_x$ of size $v_n$ and $v_x$, which marginally are given by Beta distributions: $W_n \sim \hbox{Beta}( v_n, m + 1 - v_n )$ and $W_x \sim \hbox{Beta}( v_x, m + 1 - v_x )$, where $m$ is the total number of subjects.  In words, $P$, the evidence for the hypothesis, is given by the probability that both the internal interval $W_n$ is greater than the lower quantile $q_l$ and that the external interval $W_x$ is less than the higher quantile $q_u$.  $Q$, the evidence against the hypothesis, is given by the probability that either the internal interval is too large or that the external interval is too small.  The remaining evidence is ambiguous with respect to the hypothesis, so it is assigned to $R$, ``don't know''.

In the presence of LTFs, we must consider that the actual number of failures between two time points $t_j < t_k$ is potentially unknown, but is bounded (by $d_{j,k}$ and $e_{j,k}$ as described above).  Thus where in the above formulae we used the number of failures directly from the $C$ matrix, in general we must use the appropriate upper or lower bound.  If we define the ``maximum internal interval count'' $v_n^u$ as $e_{k_{t_l}^a,k_{t_u}^b}$ and the ``minimum internal interval count'' $v_n^l$ as $d_{k_{t_l}^a,k_{t_u}^b}$, and likewise define the ``maximum external interval count'' $v_x^u$ as $e_{k_{t_l}^b,k_{t_u}^a}$ and the ``minimum external interval count'' $v_x^l$ as $d_{k_{t_l}^b,k_{t_u}^a}$, then we can in general compute the evidence $Q$ against the hypothesis as the sum of the probablities that the maximum internal interval $W_n^u$ is less than the lower quantile $q_l$ and that the minimum external interval $W_x^l$ is greater than the upper quantile $q_u$.

We compute the evidence $P$ for the hypothesis as the probability that (simultaneously) both the minimum internal interval $W_n^l$ is greater than the lower quantile $q_l$ and the maximum external interval $W_x^u$ is less than the upper quantile $q_u$.  This is most readily calculated via its complement, since this is just $1 - ( P(W_n^l \leq q_l) + P(W_x^u \geq q_u) )$.  Since these are mutually exclusive events, the probabilities may be computed separately, and each is computed simply via the Beta CDF.

\begin{figure}[ht]
%\centering
\subfigure[Example of uniform draws that support the hypothesis.]{
\includegraphics[scale=.4]{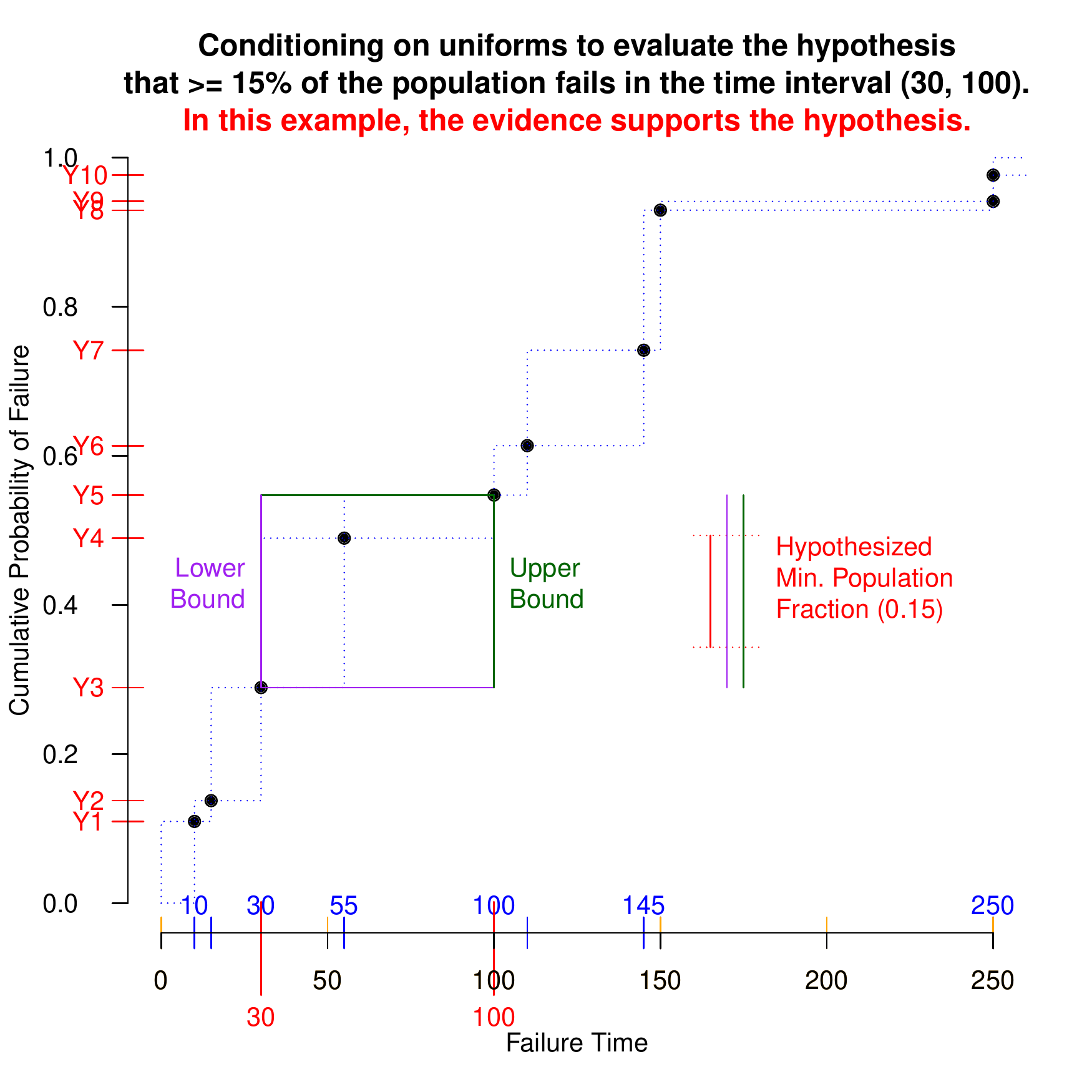}
\label{fig:onesample_obsAssertionTimes_for}
}
\subfigure[Example of uniform draws that contradict the hypothesis.]{
\includegraphics[scale=.4]{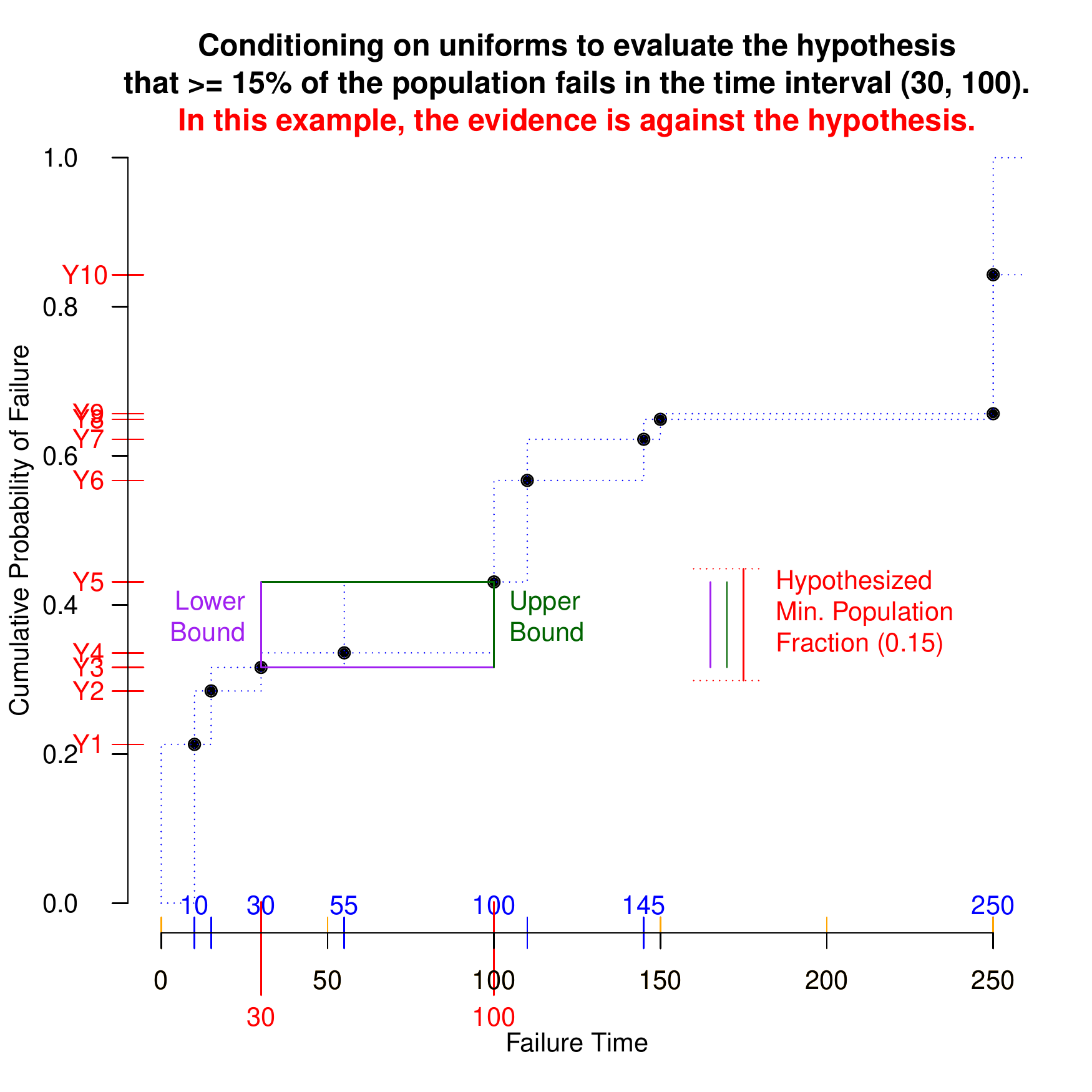}
\label{fig:onesample_obsAssertionTimes_against}
}
\label{fig:onesample_obsAssertionTimes}
\caption[]{Examples of conditioning on uniform samples to make nonparametric inferences about a CDF in the absence of missing data: evidence for \subref{fig:onesample_obsAssertionTimes_for} and against \subref{fig:onesample_obsAssertionTimes_against} the hypothesis that at least 15\% of the population fails in the time interval (30, 100).  Note that because the times 30 and 100 are observed, the evidence is unambiguous: the lower and upper bounds (on the population fraction that fail in that interval) coincide.}
\end{figure}

\begin{figure}[ht]
%\centering
\subfigure[Example of uniform draws that support the hypothesis.]{
\includegraphics[scale=.4]{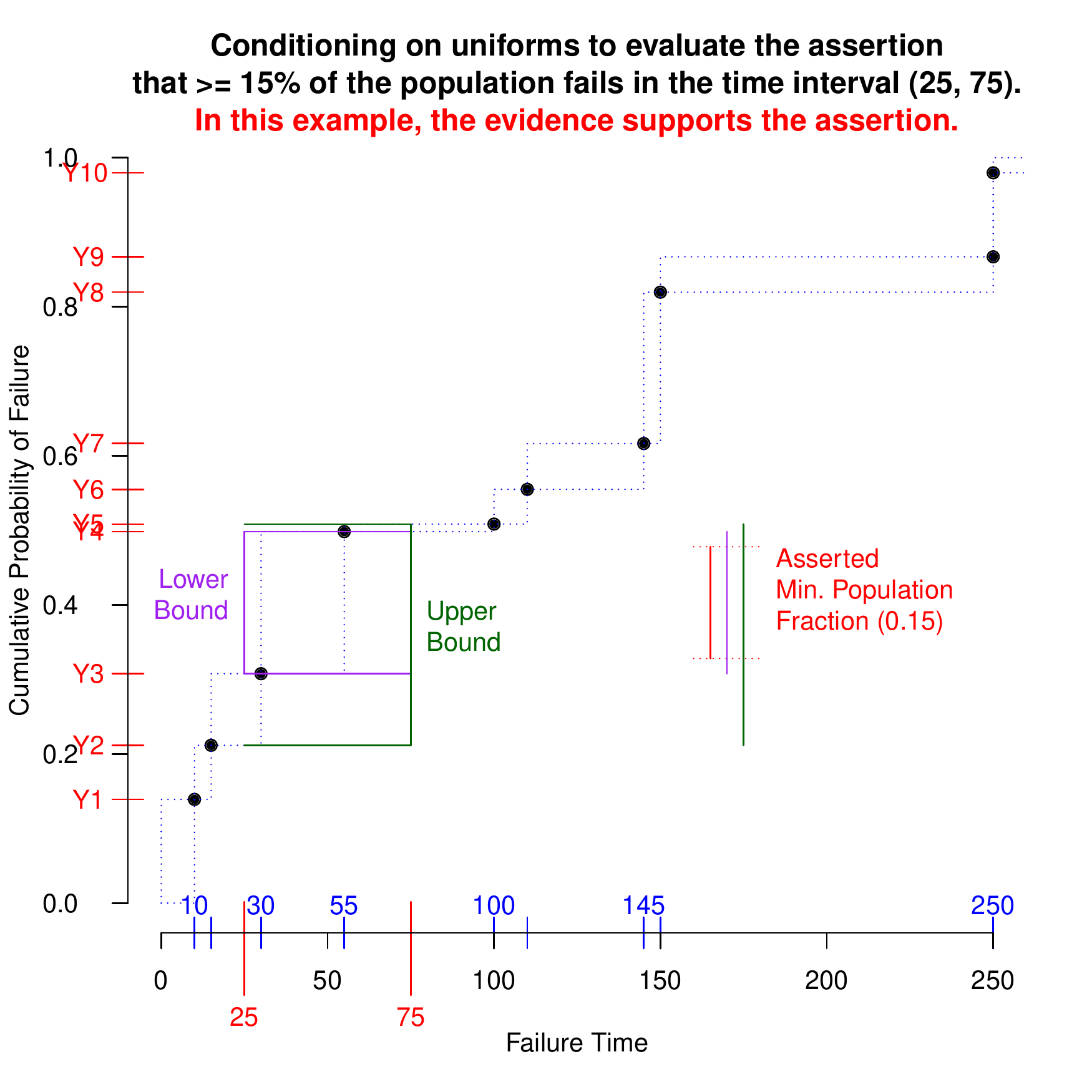}
\label{fig:onesample_for}
}
\subfigure[Example of uniform draws that contradict the hypothesis.]{
\includegraphics[scale=.4]{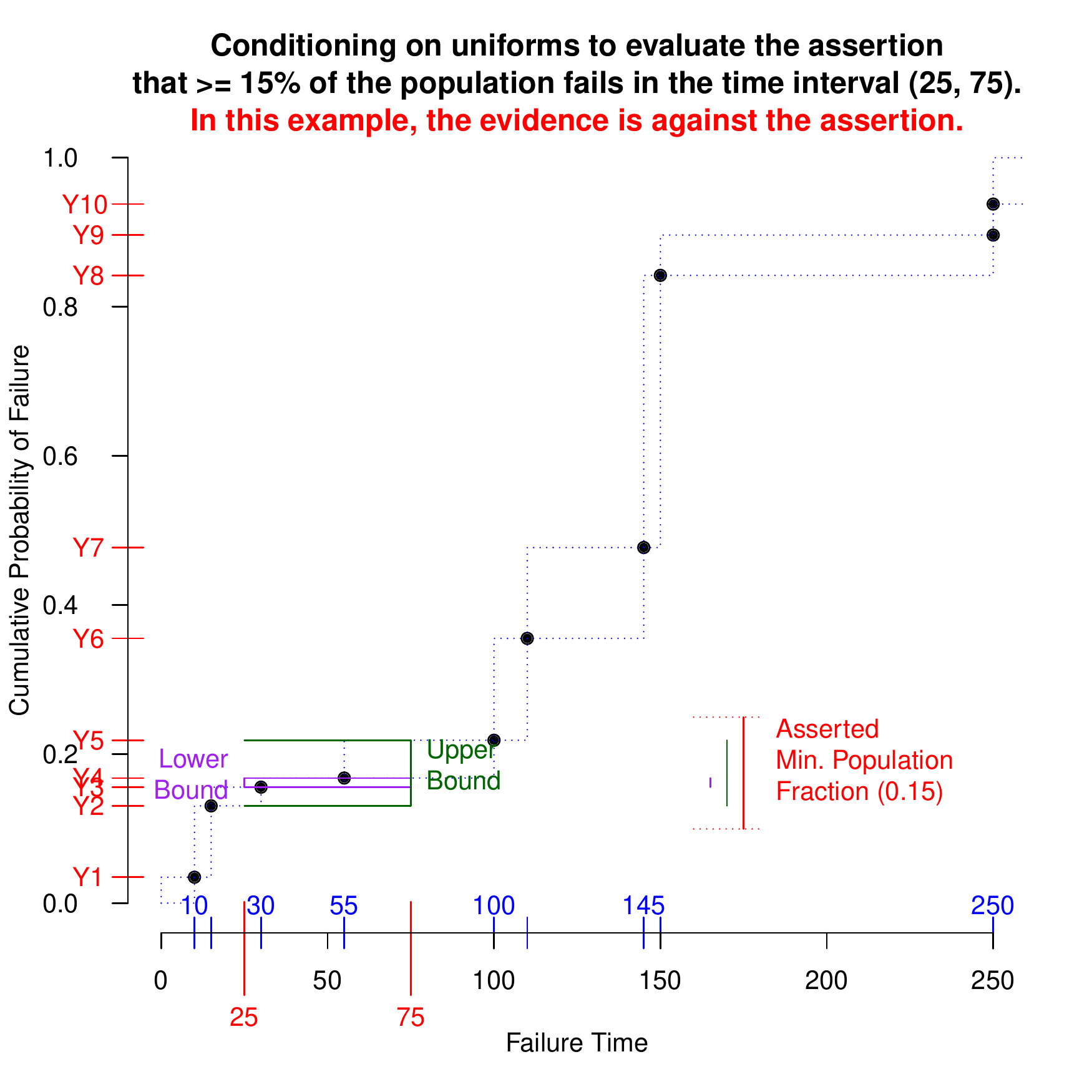}
\label{fig:onesample_against}
}
\subfigure[Example of uniform draws that are ambiguous with respect to the hypothesis.]{
\includegraphics[scale=.4]{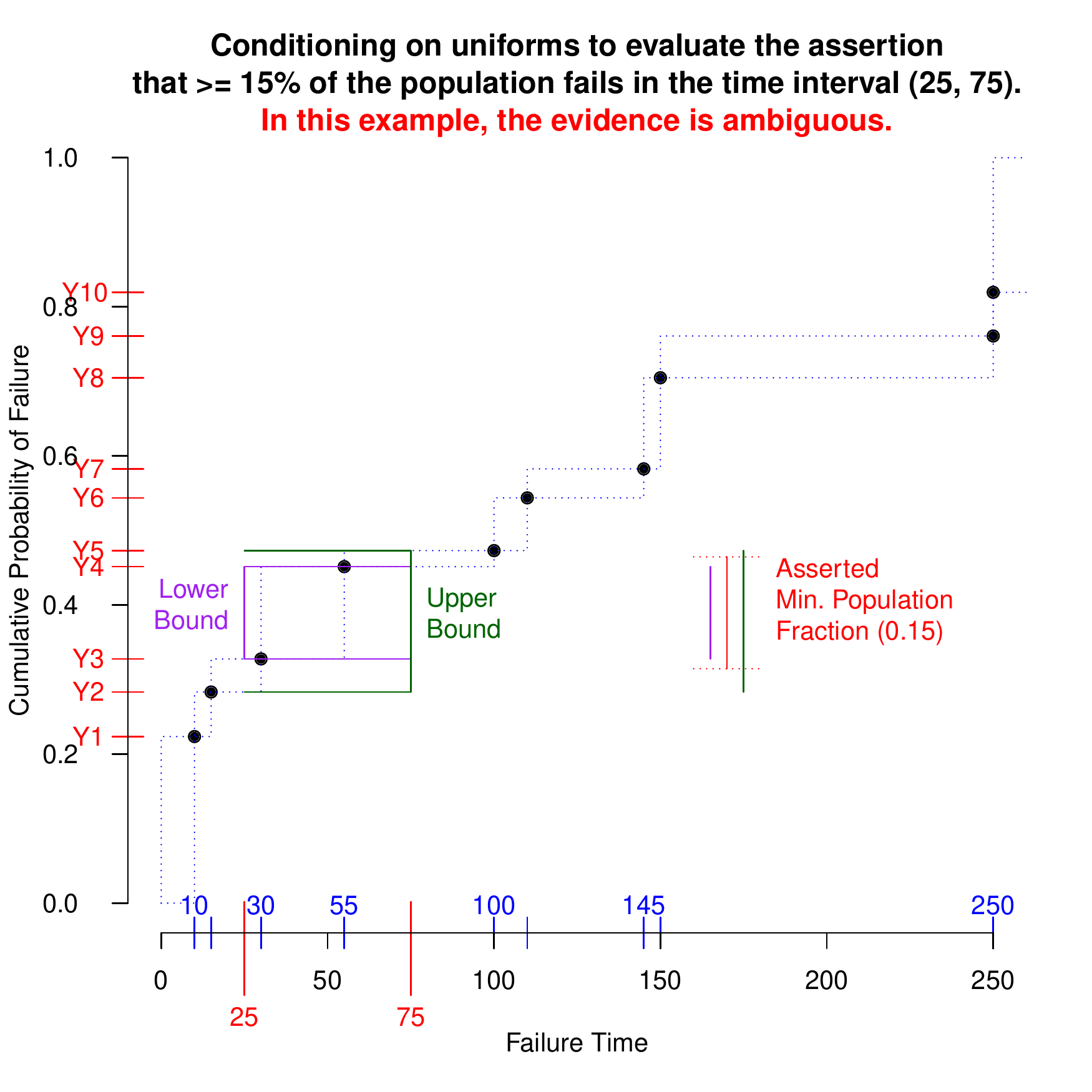}
\label{fig:onesample_ambiguous}
}
\label{fig:onesample}
\caption[]{Examples of conditioning on uniform samples to make nonparametric inferences about a CDF in the absence of missing data: evidence for \subref{fig:onesample_for}, against \subref{fig:onesample_against}, and ambiguous \subref{fig:onesample_ambiguous} with respect to the hypothesis that at least 15\% of the population fails in the time interval (25, 75).  Note that because these times (25 and 75) fall between observed times, there is ambiguity about how many failures occur, leading to a difference between the lower bound and the upper bound on the population fraction failing in that interval, which is reflected in the possibility of evidence that is ambiguous with respect to the hypothesis.}
\end{figure}

\begin{figure}[ht]
%\centering
\subfigure[Example of uniform draws that support the hypothesis.]{
\includegraphics[scale=.4]{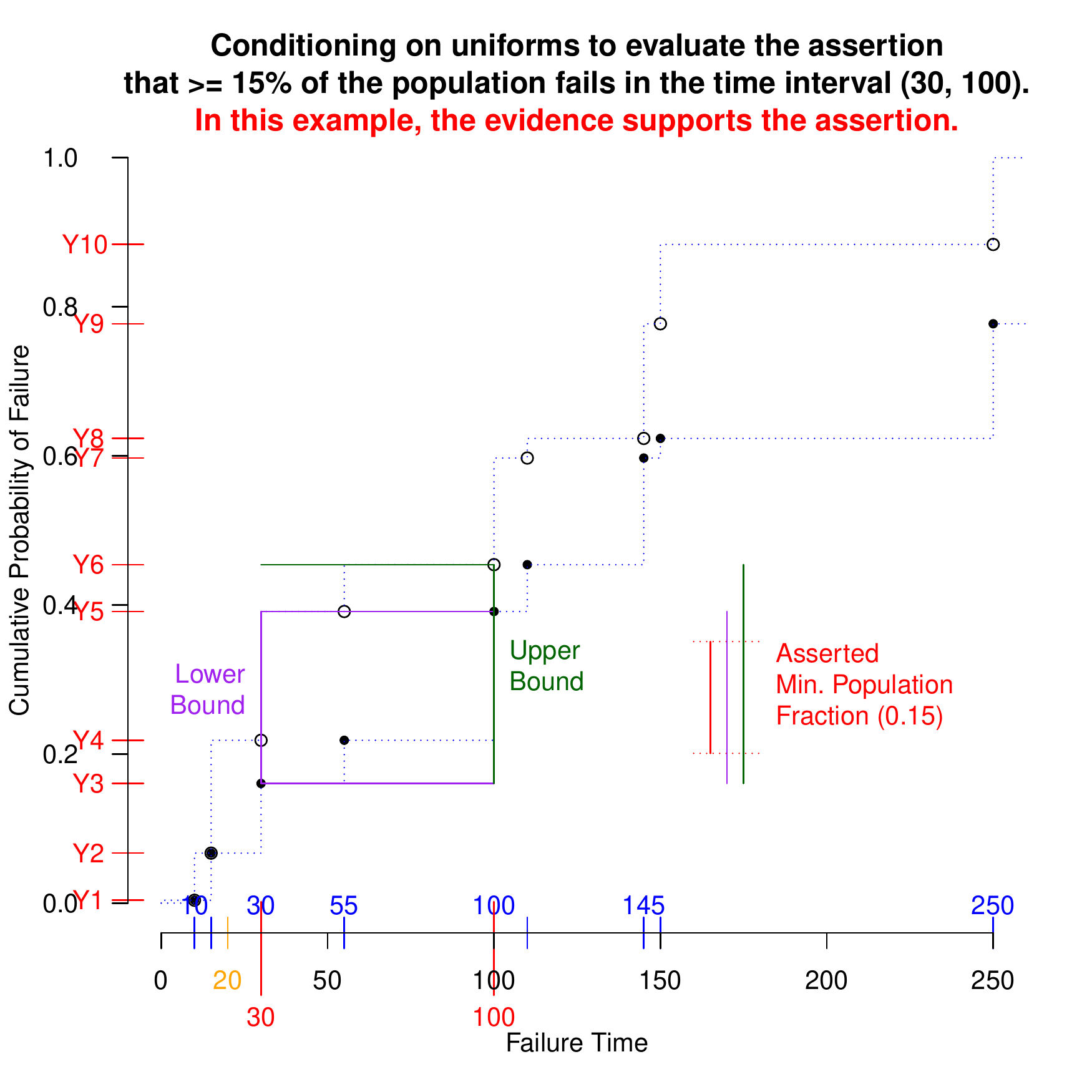}
\label{fig:onesample_obsAssertionTimes_withEarlyLTF_for}
}
\subfigure[Example of uniform draws that contradict the hypothesis.]{
\includegraphics[scale=.4]{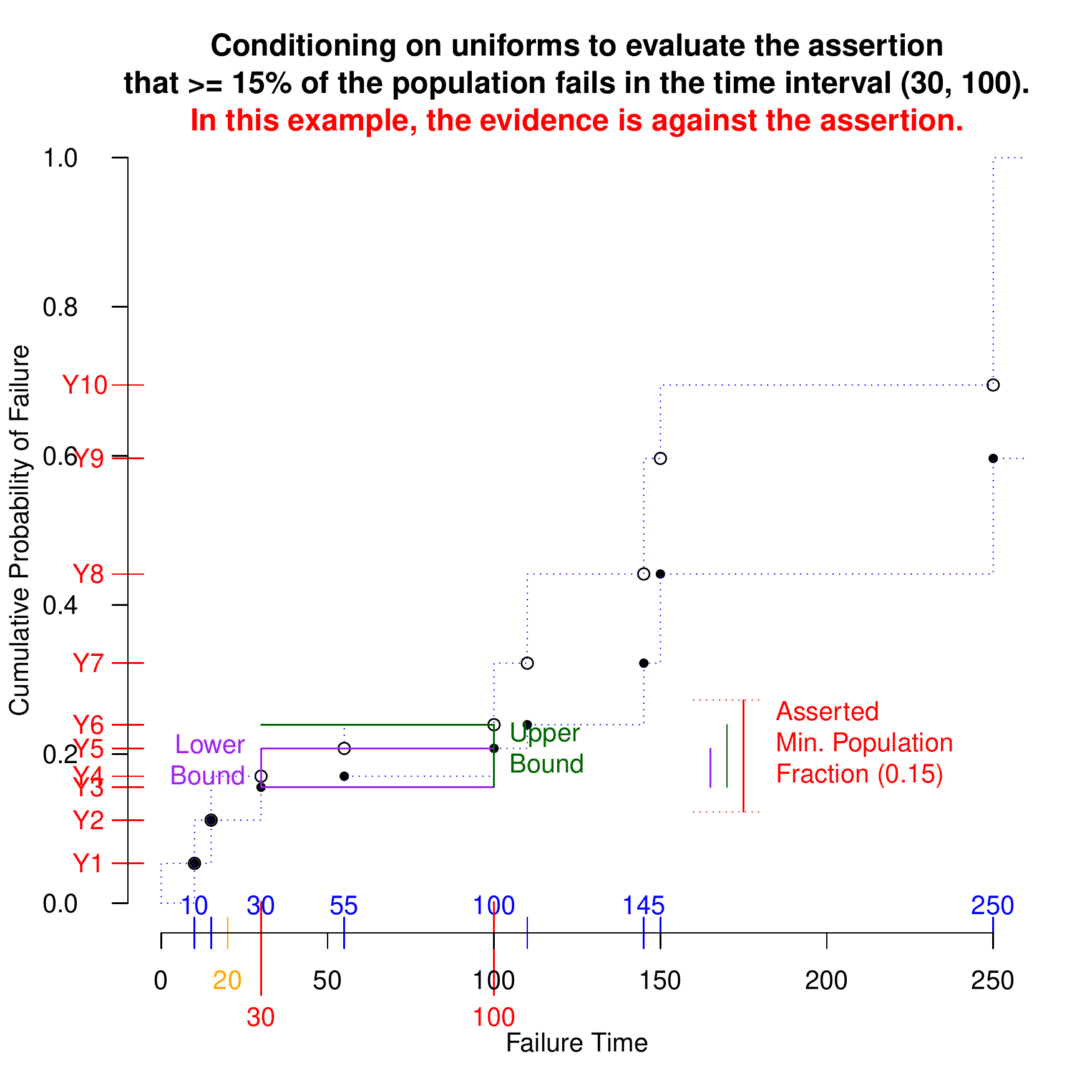}
\label{fig:onesample_obsAssertionTimes_withEarlyLTF_against}
}
\subfigure[Example of uniform draws that are ambiguous with respect to the hypothesis.]{
\includegraphics[scale=.4]{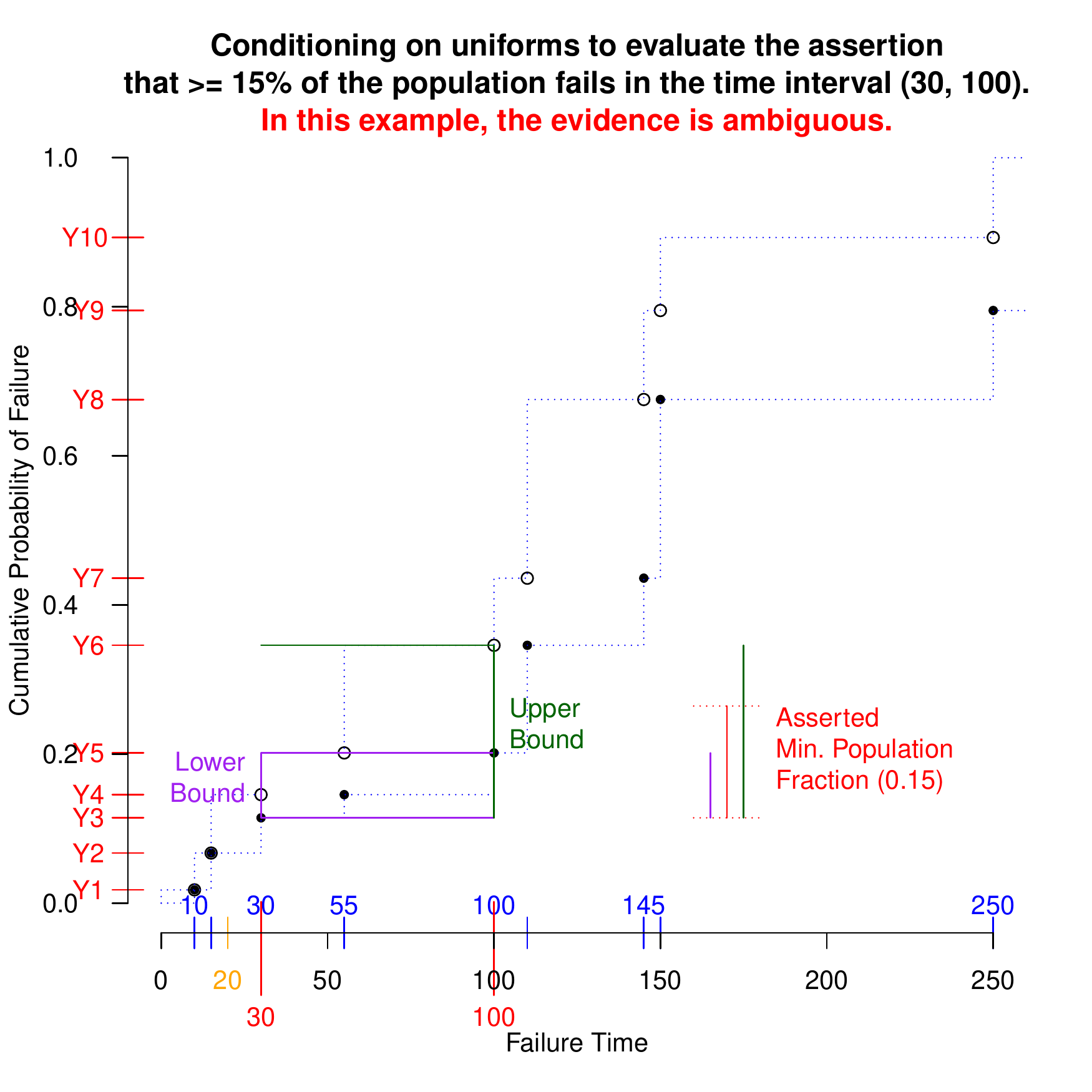}
\label{fig:onesample_obsAssertionTimes_withEarlyLTF_ambiguous}
}
\label{fig:onesample_obsAssertionTimes_withEarlyLTF}
\caption[]{Examples of conditioning on uniform samples to make nonparametric inferences about a CDF in the absence of missing data: evidence for \subref{fig:onesample_obsAssertionTimes_withEarlyLTF_for}, against \subref{fig:onesample_obsAssertionTimes_withEarlyLTF_against}, and ambiguous \subref{fig:onesample_obsAssertionTimes_withEarlyLTF_ambiguous} with respect to the hypothesis that at least 15\% of the population fails in the time interval (30, 100).  Here, ambiguity is introduced by the subject that was lost-to-followup at time 20, who may or may not have failed in the interval (30, 100).}
\end{figure}

\begin{figure}[ht]
%\centering
\subfigure[Example of uniform draws that support the hypothesis.]{
\includegraphics[scale=.4]{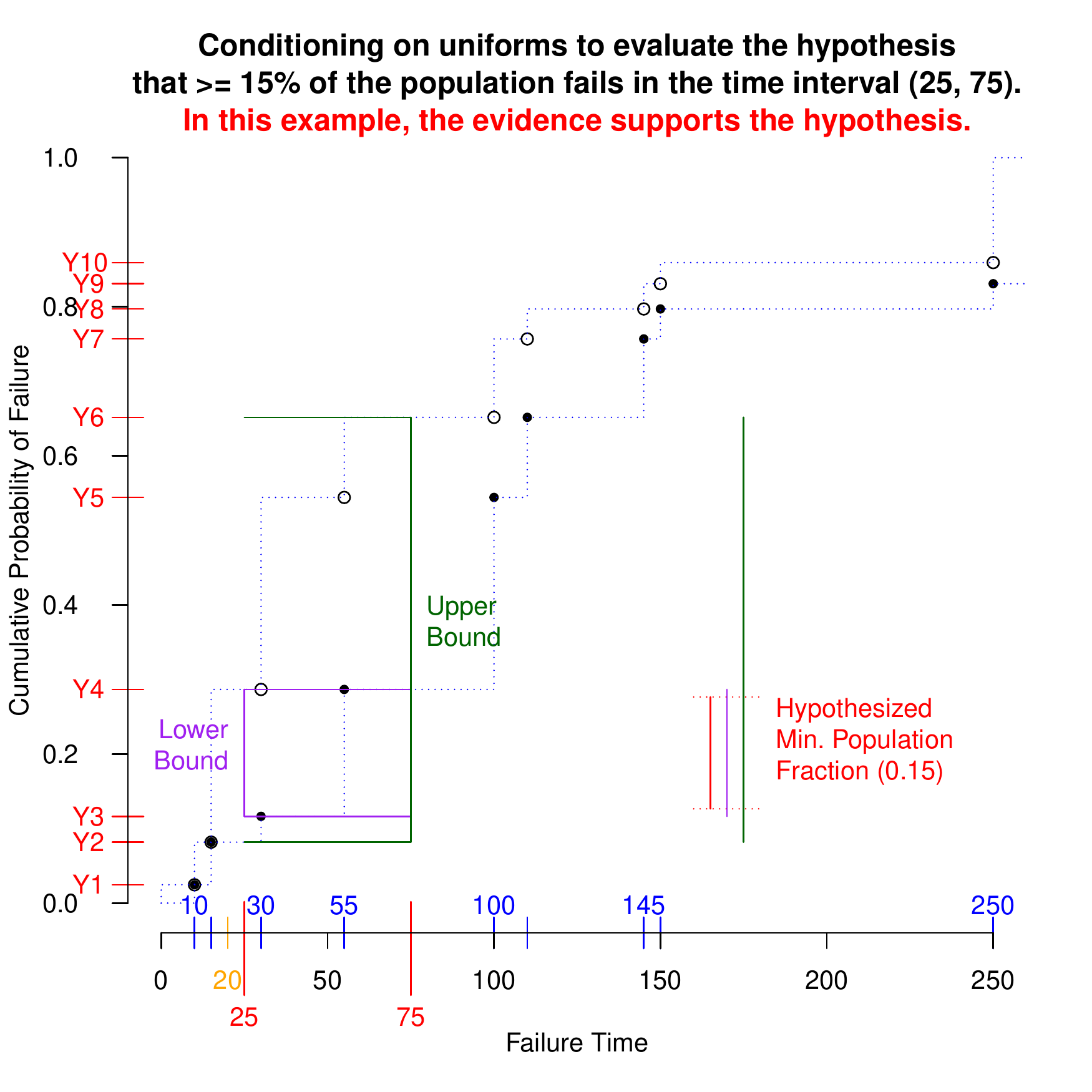}
\label{fig:onesample_withEarlyLTF_for}
}
\subfigure[Example of uniform draws that contradict the hypothesis.]{
\includegraphics[scale=.4]{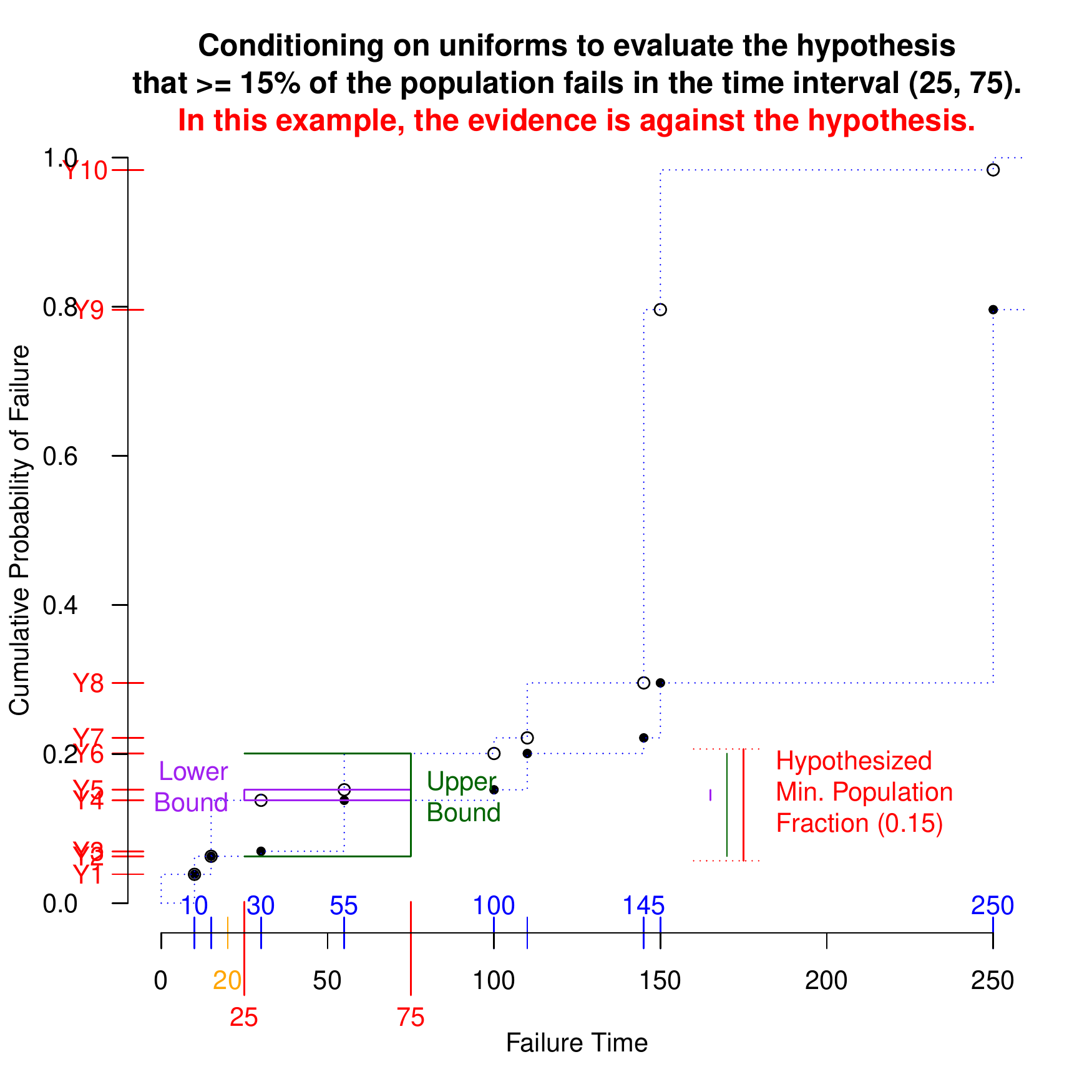}
\label{fig:onesample_withEarlyLTF_against}
}
\subfigure[Example of uniform draws that are ambiguous with respect to the hypothesis.]{
\includegraphics[scale=.4]{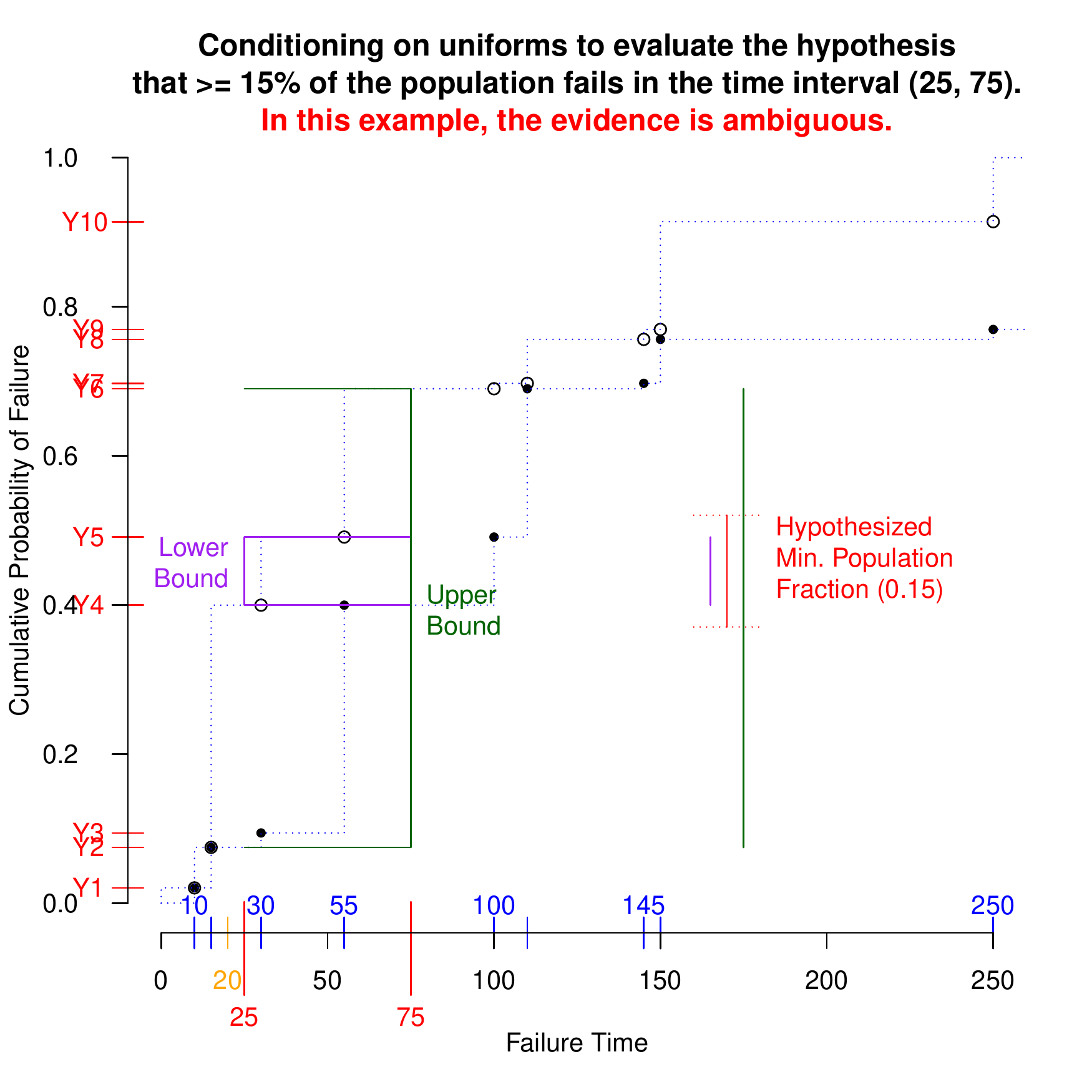}
\label{fig:onesample_withEarlyLTF_ambiguous}
}
\label{fig:onesample_withEarlyLTF}
\caption[]{Examples of conditioning on uniform samples to make nonparametric inferences about a CDF in the absence of missing data: evidence for \subref{fig:onesample_withEarlyLTF_for}, against \subref{fig:onesample_withEarlyLTF_against}, and ambiguous \subref{fig:onesample_withEarlyLTF_ambiguous} with respect to the hypothesis that at least 15\% of the population fails in the time interval (25, 75).  Here, ambiguity is introduced both by the subject that was lost-to-followup at time 20, who may or may not have failed in the interval (25, 75), and because these times (25 and 75) fall between observed times.}
\end{figure}

%% Note that if the data were observed at discrete intervals, it may be that a recorded failure or LTF occured sometime between that time point and the previous time point, and this should be accounted for in the calculation.

%\section{Nonparametric DS Estimation of VE}
%% [ERE I AM: the idea is that the ratio r_v/r_p is bounded below by the smallest r_v divided by the largest r_p, and bounded above by the largest r_v divided by the smallest r_p..]

%\bibliography{NonparametricDSHazard}

\end{document}